\documentclass[aps,twocolumn,groupedaddress, showpacs,prb]{revtex4}
\usepackage{graphicx}%

\begin{document}
\bibliographystyle{apsrev}


\title{Information physics: From energy to codes}


\author{P. Fraundorf}
\affiliation{Physics \& Astronomy, U. Missouri-StL (63121), St. Louis, MO, USA}
\email[]{pfraundorf@umsl.edu}


\date{\today}

\begin{abstract}

We illustrate in terms familiar to modern day science students that: 
(i) an {\em uncertainty slope} mechanism underlies the usefulness of temperature 
via it's reciprocal, which is incidentally around 42 [nats/eV] at the freezing 
point of water; 
(ii) {\em energy over kT} and {\em differential heat capacity} are 
``multiplicity exponents'', i.e. the bits of state information lost to the 
environment outside a system per 2-fold increase in energy and temperature 
respectively; 
(iii) even awaiting description of ``the dice'', gambling theory {\em gives 
form to} the laws of thermodynamics, availability minimization, and {\em net 
surprisals} for measuring finite distances from equilibrium, information 
content differences, and complexity; 
(iv) heat and information engine properties underlie the biological distinction 
between autotrophs and heterotrophs, and life's ongoing symbioses between 
{\em steady-state excitations} and {\em replicable codes}; and 
(v) {\em mutual information resources} (i.e. correlations between 
structures e.g. a phenomenon and it's explanation, or an organism and 
it's niche) within and across six boundary types (ranging from the 
edges of molecules to the gap between cultures) are delocalized 
physical structures whose development is a big part of the natural 
history of invention.  These tools might offer a physical framework to 
students of the code-based sciences when considering such disparate 
(and sometimes competing) issues as conservation of available work 
and the nurturing of genetic or memetic diversity. 

\end{abstract}
\maketitle

\tableofcontents
\section{Introduction}
\label{sec:Intro}

The following are part of an evolving collection of notes drawn in part 
from lecture notes taken as a student, and in part based on experiences 
teaching statistical, modern, and introductory physics.  The idea 
underlying the collection is that information theory since the days
of Shannon\cite{ShannonWeaver49, Jaynes57a, Jaynes57b} sees entropy and other thermodynamic
concepts as nothing more than tools for applying gambling theory (i.e.
statistical inference) to physical systems with large numbers of similar
and/or identical constituents. This paradigm shift\cite{Kuhn70} has already
worked its way into many advanced\cite{Grandy87, Plischke89, Garrod95} and senior
undergraduate\cite{Reif65, Katz67, Girifalco73, Kittel80p, Stowe84p, 
Baierlein99, Schroeder00} textbooks on
statistical physics. The deeper understanding and wider application, as well
as the simplifications\cite{Castle65}, that it affords to the introductory
physics student are, with few exceptions \cite{Moore98}, not yet available 
in texts. The objective here is simply
to collect some of the snapshots offered by an information theory view,
along with the calculation details and references that underlie them, for
the benefit of teachers (as well for authors as markets develop for texts 
which put these insights to use).

\section{How hot works}
\label{sec:HotWorks}

When you first heard it applied in the context of painful experience as a child, you likely gained 
appreciation for the meaning of  ``hot'' without understanding the mechanisms behind it's reputation.  
Our job here is to show you that hot, as bizarre as this may sound, means ``low uncertainty slope 
for energy exchange''.  This is an assertion that draws from the wide applicability of such slopes 
in gambling theory, which predicts for example that conserved quantities [verify that entropy's concavity 
is automatic] will likely flow from low to high slope systems when given the chance.  When the number 
of opportunities for random energy exhange are numerous, such predictions are highly accurate, 
and may be realized on so rapid a time scale that significant energy transfer occurs before your body 
has time to react and avoid damage.  When energy is the conserved quantity, the uncertainty slope is 
called reciprocal temperature or {\em coldness} $\frac{1}{k T}$.  It approaches infinity at absolute 
zero, and goes from 39 to 42 [nats/eV] as temperature decreases from room temperature to the 
freezing point of water.  Here nats is a unit of information-uncertainty defined by $\# choices = e^{\# nats}$ 
just as bits are defined by $\# choices = 2^{\# bits}$.  LASERs operate by taking advantage of 
inverted population (i.e. {\em negative} uncertainty slope) states to deliver energy most anywhere. 

\subsection{Familiar relationships} 

The systems of thermal physics traditionally involve molecules.  Hence we 
first recall how to convert between molecules N and moles n using the gas
constant $R = 8.31$[J/(mole K)].  Since R is a product of Avogadro's number 
${{{\aleph }_A}}$ (essentially the number of atomic mass units in a gram) 
and Boltzmann's constant ${{k_B}=1.38\times {{10}^{-23}}}$[J/K], one can 
write...

\begin{equation}
{N {k_B} = (N/{{\aleph }_A})({{\aleph }_A}{k_B}) = n R}
\label{Avogadro}
\end{equation}

In what follows, we will use be sticking mainly with the left 
hand side of this equation (i.e. the molecular rather than 
the macroscopic point of view), and be using a quantity k to 
determine the units used for temperature T.  In the particular 
case when k is chosen to be ${{k_B}}$, the temperature will be 
in historical units (e.g. [Kelvins]).  When $k=1$, or when we 
equivalently consider the quantity $k T$ rather than $T$ as 
the temperature, then we will say that temperature is in 
``natural units''.  Below, we show that in natural units 
temperature may be expressed in Joules (or electron volts) per 
nat of mutual information about an object's state lost to 
the world around.

Before examining this in more detail, let's consider a couple of 
useful elementary thermodynamic relationships: ``equipartition'' 
and ``the ideal gas equation of state''.
\begin{equation}
\text{Equipartition:  } U = \frac{\nu N}{2} k T = \frac{\nu n}{2} R T
\label{Equipartition}
\end{equation}
Here $\nu$ is often called the number of {\em degrees of freedom}, or 
modes of thermal energy storage, per molecule.  The equation 
relates extensive quantity $U$, the amount of randomly-distributed 
mechanical (kinetic and potential) energy in a gas or solid, to an 
intensive quantity: its absolute temperature $T$.  We show later how 
this relation arises from the equation for a quadratic system's number 
of accessible states.  Likewise, the equation of state for an ideal 
gas below follows from the assumption that each molecule in an ideal 
gas has a volume of $V$ in which to ``get lost''.

\begin{equation}
\text{Ideal Gas:  } P V = N k T = n R T
\label{IdealGas}
\end{equation}
The above equation thus relates extensive quantity {\em volume} $V$  
to intensive quantities: {\em absolute temperature} $T$ and 
{\em pressure} $P$.

Using these two equations, show that energy and temperature are quite 
different by proving to yourself that when you build a fire in an igloo, the
total thermal kinetic energy of the air inside is unchanged \cite{HRW97}!  
(Hint:  This is true even though the temperature of the air goes up.) 

\subsection{Law zero with teeth}

To examine the way that thermal physics can give birth to such relations, 
a useful concept is the {\em multiplicity} or ``number of accessible states'' 
$\Omega$.  Since for macroscopic systems this is often an unimaginably huge 
number (on the order of $e^{\aleph_A}$), one commonly deals 
with its logarithm the uncertainty or ``entropy'' $S = k \ln \Omega$.  (Look 
for more on the connection between uncertainties which depend on one's 
frame of reference, and physical entropy, later.)  $S$ is measured in 
information units [nats, bits, or J/K] depending on whether $k$ is chosen 
to be \{1, $\frac{1}{\ln 2}$, or $k_B$ \} respectively.  Knowing the dependence 
of multiplicity nd hence $S$ on any conserved quantity $X$ (like energy, volume, 
or number of particles), shared randomly between two systems, allows one to 
``guess'' how $X$ is likely to end up distributed between the two systems.  
One simply chooses that sharing of $X$ which can happen in the largest 
number of ways, a mathematical exercise (try doing it yourself!) which for 
reasonable functions predicts that systems will most likely adopt subsystem 
$X$-values for which subsystem uncertainty slopes $\frac{dS}{dX}$ are equal, i.e.  
\begin{equation}
\text{X equilibrated} \Rightarrow S_{tot} \text{maxmized} \Rightarrow \text{all} \frac{\partial S_i}{\partial X} \text{equal.}
\label{Law0}
\end{equation}

\subsubsection{Energy \& equipartition}

This simple assertion yields some powerful results.  Consider first 
the large class of macroscopic systems which can be classified as 
``approximately quadratic in thermal energy''.  For these we can write 
$\Omega \propto U^{\frac{\nu N}{2}}$, where as above $N$ is the number 
of molecules and $\nu$ is the number of degrees freedom per molecule.  
Such systems include low density gases, metals near room temperature, 
and many other macroscopic systems at least in some part of their 
temperature range.  Using $c$ to denote a constant not dependent on 
energy $U$, one can then calculate uncertainty $S$ and it's first and 
second derivatives:  
\begin{equation}
\Omega \propto U^{\frac{\nu N}{2}} \Rightarrow \frac{S}{k} = \frac{\nu N}{2} \ln U + c\text{.}
\label{X1}
\end{equation}
The first derivative says that the energy uncertainty slope of such 
systems, a quantity predicted to ``become the same for all subsystems 
allowed to equilibrate in thermal contact'', is 
$\frac{\partial S}{\partial U} = \frac{\nu N k}{2 U}$.  This quantity 
is in historical parlance known as reciprocal temperature, i.e. as 
$\frac{1}{T}$.  One can thus solve this equation for energy to get 
the equipartition relation above: $U=\frac{\nu N}{2}$.  

The second energy derivative of uncertainty is the negative quantity 
$\frac{\partial^2 S}{\partial U^2} = - \frac{\nu N k}{2 U^2}$.  Hence systems 
with greater energy have lower uncertainty slope.  As a result, energy flow 
during thermal equilibration goes from systems of lower to higher 
uncertainty slope, and equivalently from higher to lower temperature.   
This rate of uncertainty increase per unit energy gain (also called ``coldness'') 
thus behaves like a kind of hunger for thermal energy, just as gas pressure 
(below) can be seen as an appetite to acquire volume.  By comparison, hot 
objects are like reservoirs of excess thermal energy which has limited room 
to play.  Hence the energy uncertainty slope (about 42 nats/eV at room 
temperature, running to infinity as one approaches absolute zero) effectively 
drives the random flow of heat.  The second derivative calculation above (by 
taking a square root) also allows one to estimate the size of observed 
temperature (or energy) fluctuations, as will be shown more quantitatively 
later.

\subsubsection{Volume \& ideal gas laws}

A system that has a simple volume dependence for the number of accessible 
states is the ideal gas.  If the gas has sufficiently low densities that gas 
molecules seldom encounter one another, then the number of places any 
particular gas molecule may occupy is likely proportional to the volume $V$ to 
which the gas is confined.  Moreover, the independence of molecules in this 
low density (ideal gas) case means that the number of accessible states for 
the gas as a whole is simply proportional to the product of the number of 
states for each molecule separately, so that $\Omega \propto V^N$.  As above, 
we can then calculate uncertainty and its first and second derivatives:
\begin{equation}
\Omega \propto V^N \Rightarrow \frac{S}{k} = N \ln V + c\text{.}
\label{X1}
\end{equation}
The $1^{st}$ derivative is $\frac{\partial S}{\partial V} = \frac{N k}{V}$ 
and the $2^{nd}$ derivative is $\frac{\partial^2 S}{\partial V^2} = -\frac{N k}{V^2}$.
The negative value of the latter suggests again that volume is likely to 
spontaneously flow (when being randomly shared) from systems of lower 
uncertainty slope to systems of higher slope.  

But what is the physical meaning of {\em free expansion coefficient} 
$\gamma \equiv \frac{\partial S}{\partial V}$?  A clue 
might come from thinking of it as a product of $\frac{\partial S}{\partial U}$ 
and $\frac{\partial U}{\partial V}$.  As discussed above, the former is 
normally written as $\frac{1}{T}$, while $\frac{\partial U}{\partial V}$ is 
nothing other than the change in energy per unit volume as in $Work = P dV$
or in other words a pressure.  Thus the calculation above tells us that 
$\frac{P}{T}$ (the free expansion coefficient for an ideal gas) equals 
$\frac{N k}{V}$.  Hence the ideal gas law!

This volume uncertainty slope, in natural units at standard temperature and 
pressure, is about $2.5\times 10^{19}$[nats/cc] at standard temperature and 
pressure: much less than the atomic density of around $10^{23}$ atoms per cc 
for solids.  The negative 2nd derivative predicts that for systems at the 
same temperature*, volume will ``spontaneously flow'' from systems of lower 
to higher pressure.  Put another way, high pressure systems will expand at 
the expense of the low pressure neighbors, something that is quite consistent 
with observation.

* A thermally-insulating barrier between two systems which allows ``totally 
random sharing of volume'' is difficult to imagine.  Easy to imagine is a 
rigid but mobile partition, dividing a closed cylinder into two 
gas-tight halves.  In this case, gases on opposite sides will adjust P to a 
common value on both sides of the barrier, thus establishing mechanical 
(momentum transfer) equilibrium with unequal densities and temperatures.  
The higher temperature (lower density) side will then experience fewer, 
albeit higher energy, collisions.  These will eventually result in thermal 
equilibration by differential energy transfer, even if we have to think 
of the wall as a single giant molecule with one degree of freedom, whose 
own average kinetic energy will ``communicate'' uncertainty slope differences 
between sides.

\subsubsection{Particles \& mass action}

The random sharing of particles (for example in a reaction) also gains it's 
sense of direction from the $0^{th}$ Law of Thermodynamics described here.  
First, determine how accessible states depends on the number of particles.  
Taking derivatives of uncertainty with respect to particle number for an 
ideal gas, one finds that $\frac{\partial S}{\partial N}$ (also known as 
chemical affinity $\alpha \equiv \frac{-\mu}{k T}$) approaches 
$\ln \frac{Q}{N/V}$, where ``quantum concentration'' Q is the number of 
particles per unit volume allowed by thermal limits on particle movement.  
Here $\frac{Q}{N/V}$ is effectively the number of available non-interacting 
quantum states per particle.  As particle number density $n  \equiv  N/V$ 
increases toward Q, affinity $\alpha$ (near 16[nats/particle] for Argon gas 
at standard temperature and pressure) decreases toward 1, and ideality is 
lost.  Ratios between $Q_i$ values in gas reactions, for the various 
components i of a reaction, yield an equilibrium constant that allows one to 
predict ratios between resulting concentrations $n_i$. 

For example, if we consider the reaction $A_2+2B\leftrightarrow 2AB$, we expect
equilibrium when the affinities of reactants on both sides are equal, i.e. when 
\begin{equation}
\alpha _{A_2}+2\alpha _B=2\alpha _{AB}\text{.} 
\label{AffinityBalance}
\end{equation}
Hence {\em the equilibrium constant} is 
\begin{equation}
K\equiv \frac{n_{AB}^2}{n_{A_2}n_B^2}=\frac{Q_{AB}^2}{Q_{A_2}Q_B^2}\text{,} 
\label{EqConst}
\end{equation}
where the middle term depends only on reactant concentrations, while the last
term is a function of experimental conditions e.g. for a monatomic ideal gas 
$Q = \left(2\pi m k T/h^2\right) ^{3/2}$ where $m$ is each atom's 
mass and $h$ is Planck's constant. Thus the behavior of the concentration 
balance as a function of temperature may be predicted. 

\section{The first and second laws}
\label{sec:Laws12}

  In addition to the $0^{th}$ Law and some state equations, one can 
get the {$1^{st}$ and $2^{nd}$ Laws of 
Thermodynamics by combining gambling theory with conservation of energy and 
other shared variables.  We first illustrate with some heuristic arguments, 
and then present some more general results with help from the maximum 
entropy formalism.

For an {\em isolated} system (one cut off from ourselves and the rest of 
the world), the first and second laws are intuitive.  Conservation of energy 
$U$ requires that 
\begin{equation}
dU_{tot}/dt=0 \text{,}
\label{Law1}
\end{equation}
and intuition suggests that our uncertainty 
$S = k \ln[\Omega]$ about the microscopic state of a system (while we're cut 
off from it) is unlikely to decrease with time.  Although this may sound like 
it makes our presence as observers crucial to the time evolution of a system 
from which we are isolated, it does not.  As we show below, it is instead 
equivalent to saying that the mutual information 
between isolated systems is not likely to increase with time.  Thus 
it is a prediction about the behavior of the larger system of which these 
subsystems are a part.  It does have a direct impact on how our assertions about 
that system's state as a function of time will correlate with what we find, 
should we decide to terminate the isolation at some point and look inside.

The classical example of such irreversible change is the free expansion of 
a gas confined to one half of an evacuated volume, upon failure of a partition 
dividing that volume in half.  If the gas is ideal, in fact, the number of 
accessible states per particle doubles so that the entropy increase is one bit 
per particle.  Such isolated system entropy increases are called irreversible, 
and hence we can write:
\begin{equation}
\frac{dS_{tot}}{dt} = \frac{\delta S_{irr}}{dt} \geq 0 \text{.}
\label{Law2}
\end{equation}
Equation \ref{Law1} is rigorous within limits of the energy-time uncertainty 
principle, while equation \ref{Law2} is only a probabilistic assertion, 
albeit one often backed up by excellent statistics!

For a system allowed to share thermal energy U, volume V, and particles N with 
its environment, the change of entropy can be written:
\begin{equation}
dS=\left( \frac{\partial S}{\partial U}\right)_{VN} dU + 
\left( \frac{\partial S}{\partial V}\right)_{NU} dV + 
\left( \frac{\partial S}{\partial N}\right)_{UV} dN + \delta S_{irr} \text{.}
\label{EntropyOpen}
\end{equation}
The first term in parenthesis is $1/T$, the second $P/T$, and the third $-\mu /T$ from our 
statistical definitions of the intensive variables.  If we solve this for dU, while 
defining as flows of heat $\delta Q$ those energy changes NOT associated with changes 
in a specific extensive variable (like V or N), we get the most common ``open system'' 
version of the First Law,
\begin{equation}
dU = \delta Q_{in} - \delta W_{out} \text{.}
\label{Law1open}
\end{equation}
Here $\delta W \equiv P dV - \mu dN$ denotes work done by the system on its 
external environment as it gains volume or loses particles.  The resulting equality 
of $\delta Q$ with $T(dS-\delta S_{irr})$, rearranged, yields the open 
system form for the Second Law:
\begin{equation}
dS = \frac{\delta Q_{in}}{T} + \delta{S_{irr}} \text{, where} \frac{\delta S_{irr}}{dt} \geq 0\text{.}
\label{Law2open}
\end{equation}
Here of course $W_{out}$, $Q_{in}$ and $S_{irr}$ are defined only in the context 
of their respective pathways for energy or entropy change, and are not 
themselves functions of the state of the system at all.  The use here 
of $\delta$, instead of $d$, to represent their differentials is thus 
because those differentials are mathematically {\em inexact}\cite{Stowe84p}.

Note that in the process we have also shown, for reversible changes, 
i.e. when $\delta S_{irr}=0$, that $P=-(\partial U/\partial V)_{SN}$ is a 
measure of force per unit area, and $\mu = (\partial U/\partial N)_{SV}$ is 
a measure of energy per particle.  Of course, with added terms of the same 
form equation \ref{Law2open} can accomodate many simultaneous kinds of 
work and particle exchange.

Given the 1st and 2nd Laws, along with $\Omega(U,V)$ for an ideal gas 
an its consequences, a large number of simple but interesting problems 
can be considered by introductory students {\em in detail}.  These 
include a host of gas expansion problems {e.g. isothermal, isobaric, 
isochoric, adiabatic, and free), a set of two-system problems which 
include information loss during irreversible cooling e.g. a cup of 
coffee whose initial net surprisal from subsubsection 
\ref{subsec:MaxEnt6} is $\approx 
C_v(\frac{T_c}{T_r}-1-\ln\frac{T_c}{T_r})$, attempts by Maxwell's 
Demon at reversing the heat flow 
process, and the symmetric vacuum-pump memory (or isothermal compressor) 
discussed in subsection \ref{subsec:InformationEngines}.  
Second Law limits on converting high temperature heat 
to room temperature in the presence of an external low temperature reservoir 
also yield suprising results \cite{Jaynes91}.  If the dependence of 
$\Omega$ on $N$ can be introduced, as discussed above entropies of mixing 
and chemical reaction rates may be considered as well.

\section{The maxent ``best-guess'' machine}
\label{sec:MaxEnt}

If one has information over and above the state inventory needed to 
determine $\Omega$, it can be used to modify the assignment of equal a 
priori probabilities e.g. used above when we maximized uncertainty for the 
(``microcanonical'') case in which extensive variables (like energy, volume 
and number of particles) are all considered independent variables or work 
parameters \cite{Jaynes79}.  To do this, one first writes entropy in terms 
of probabilities by defining for each probability $p_i$ a ``surprisal'' 
$s_i \equiv k \ln[1/p_i]$, in units determined by the value of $k$.  
The average value of this surprisal reduces to $S = k \ln \Omega$ when the 
$p_i$ are all equal, yielding a generalized multiplicity $\Omega \equiv e^{S/k}$.  
Note that the relationships described here will likely translate 
seamlessly into quantum mechanical applications \cite{Jaynes57b}.
 
\subsection{The problem}

Our (at first glance benign) task is to maximize average surprisal...
\begin{equation}
\frac{S}{k}=\left< \frac{s}{k} \right>=\sum _{i=1}^{\Omega }{p_i}\frac{{s_i}}{k}=\sum _{i=1}^{\Omega }{p_i}\ln (\frac{1}{{p_i}})
\label{AverageSuprisal}
\end{equation}
...subject to the normalization requirement that the probabilities add to 1, i.e. that...
\begin{equation}
\sum_{i=1}^{\Omega} p_i = 1\text{,}
\label{Normalization}
\end{equation}
...along with the ``expected average'' of $R$ constraints which take the form...
\begin{equation}
E_r=\left<e_r \right>=\sum _{i=1}^{\Omega }{p_i} e_{ri}\text{,} \forall r \in \{1,R\}\text{.} 
\label{Constraints}
\end{equation}

A simple example (useful for free electron and neutron gas models) can be thought of as that of a weighted coin which lands ``heads up'' (to be specific) six tenths of the time.  In that case, there would be $\Omega = 2$ states, and we would have $R=1$ e.g. with $e_{11} = 0$, $e_{12} = 1$, and $E_1 = 0.6$.  A more general example, that of the maxent calculation underlying the Bell curve, is presented in Appendix \ref{app:MultipleChoice}.

\subsection{The solution}

The Lagrange method of undetermined multipliers tells us that the 
solution for the $i$th of $\Omega$ probabilities is simply...
\begin{equation}
{p_i}=\frac{1}{Z}{{e}^{-\sum _{r=1}^{R}{{\lambda }_r}{e_{{ri}}}}},\forall i\in \{1,\Omega \}\text{,}
\label{Solution}
\end{equation}
where partition function $Z$ is defined to normalize probabilities as...
\begin{equation}
Z=\sum _{i=1}^{\Omega }{{e}^{-\sum _{r=1}^{R}{{\lambda }_r}{e_{{ri}}}}}.
\label{PartitionFn}
\end{equation}
Here ${{{\lambda }_r}}$ is the Lagrange (or ``heat'') multiplier for the $r$th 
constraint, and ${{e_{{ri}}}}$ is
the value of the $r$th parameter when the system is in the $i$th accessible state.  
For example, when ${{E_r}}$ is the energy $U$, ${{{\lambda}_r}}$ is often written 
as ${\frac{1}{{kT}}}$.  Values for these multipliers can be calculated 
by substituting the two equations above back into the constraint equations, or 
from the differential relations derived below.

For example, equation \ref{PartitionFn} gives $Z=1+e^{-\lambda_1}$ for our 
coin problem, so that $p_{11} = \frac{1}{Z}$, $p_{12} = \frac{e^{-\lambda_1}}{Z}$ and 
$E_1 = \frac{1}{1+e^\lambda_1 }$.  Solving in terms of $E_1$ we get 
$\lambda_1 = \ln (\frac{1}{E_1} -1) = -.405$, $Z=2.5$, $p_{11} = 0.4$, 
$p_{12} = 0.6$, and $\lambda_1 E_1$ (a quantity which will prove useful 
later) is $-.243$.

The resulting entropy, maximized under specified constraints, is...
\begin{equation}
\frac{S}{k}=\ln [Z]+\sum_{r=1}^{R}{{\lambda }_r}{E_r}.
\label{MaxEnt}
\end{equation}
A useful quantity which has been minimized in constraint-free fashion,
by this calculation, is the ``availability in information units'' 
\begin{equation}
A \equiv {- k \ln Z} = k \sum_{r=1}^{R} {\lambda }_r E_r - S\text{.}
\label{AvailBits}
\end{equation}

For example, in the coin problem the maximized entropy is 
$S/k = \ln[Z] + \lambda_1 E_1 = 0.673$ nats, and the minimized 
availability $A/k = \lambda_1 E_1 - S = -.916$ nats.  Since constrained 
entropy maximization minimizes availability without constraints, 
assuming of course that the ${{e_{{ri}}}}$ coefficients for all values of 
$r$ and $i$ are held constant themselves (cf. page 46 of Betts and Turner), 
gambling theory most simply states that the best guess (in the absence of 
other information) is the state with minimim availability.  

Generalized availability in turn can be seen as the common numerator behind 
a range of dimensioned availabilities, with the properties of thermodynamic 
free energy (one for each variable of type $r$).  These are 
defined as...
\begin{equation}
A_r=\frac{-\ln Z}{{{\lambda }_r}}=\sum_{u=1}^{R}\frac{{{\lambda }_u}}{{{\lambda }_r}}{E_u}-\frac{S}{k {{\lambda}_r}},\forall r\in \{1,R\}.
\label{DimAvail}
\end{equation}

Standard thermodynamic applications include {\em microcanonical 
ensemble} calculations like those with which we began this note ($R\equiv 0$ so that 
$A=-S$), the {\em canonical ensemble} for systems in contact with a heat bath at fixed 
temperature (there $R=1$ and ${{E_1}}$ is energy $U$ so that 
${{{\lambda }_1}}=\frac{1}{kT}$, and 
${{A_1}}=U-TS$ is the Helmholtz free energy), the {\em pressure ensemble} ($R=2$ 
with ${{E_1}}$ is energy $U$ so that ${{{\lambda }_1}}=\frac{1}{kT}$, ${{E_2}}$ 
is volume V so that ${{{\lambda}_2}}=\frac{P}{kT}$, and ${{A_1}}=U+PV-TS$ is the 
Gibbs free energy), and the {\em grand canonical ensemble} for ``open systems'' (same 
as pressure except that ${{E_2}}=N$ so that ${{{\lambda }_2}}= \frac{-\mu}{kT}$, 
and ${{A_1}}= E - \mu N - T S$ is the grand potential).  

Note that this calculation 
requires an assignment of the $e_{ri}$ for all possible states, 
but it involves no other physical assumptions (like equilibrium 
or energy conservation).  In the sense given here, systems 
for which the recommended guess is not appropriate are systems 
about which more is known (e.g. about constraints or state 
structure) than is taken into account.  Types 
of constraints other than the averages used in equation \ref{Constraints}, 
e.g. correlation information constraints like those discussed later, are 
likely worth learning to put to use, but that is not done here. 

\subsubsection{Physics-free laws 1 \& 2}

We now begin to look at the effect of small changes.  For example,
the ${{e_{{ri}}}}$ are often not themselves constant, but depend on 
the value of a set of ``work parameters'' ${{X_m}}$, where here m 
is an integer between $1$ and $M$.  For example in the Gibb's canonical 
ensemble calculations for an ideal gas, the energies of the various 
allowed states may depend on volume $V$ or particle number $N$.  Following 
Jaynes we can define work-types for each constraint $r$ in terms of the rate 
at which ${{E_r}}$ changes with ${{X_m}}$:
\begin{widetext}
\begin{equation}
\delta W_r \equiv 
\sum _{m=1}^M {{\left(-\frac{\partial {E_r}}{\partial {X_m}}\right)}} 
\delta X_m
=-\sum _{m=1}^M {{{\delta X}}_m}\sum _{i=1}^{\Omega }
\left( {e_{{ri}}}\frac{\partial {p_i}}{\partial {X_m}}
+{p_i}\frac{\partial {e_{{ri}}}}{\partial {X_m}}\right), \forall r\in \{1,R\}
\label{WorkParams}
\end{equation}
\end{widetext}
Note that the various ``work increments'' ${{{{\delta W}}_r}}$ have the same units 
as the corresponding contraint parameters ${{E_r}}$, and unless otherwise noted 
that partials are taken under ``ensemble conditions'' (i.e. holding 
constant all unused ``control'' parameters $X_m$ and $\lambda_r$).  

It's also useful when discussing work to define the generalized enthalpies
\begin{equation}
H_r\equiv E_r + \sum _{m=1}^{M}{{\left(-\frac{\partial {E_r}}{\partial {X_m}}\right)}} X_m, \forall r\in \{1,R\},
\label{Enthalpies}
\end{equation}
For example, in the canonical ensemble case mentioned above, with volume 
V the only allowed work parameter, equation \ref{WorkParams} becomes 
simply $\delta W_1 = P \delta V$, and equation \ref{Enthalpies} becomes 
$H_1 = U + P V$.

In equation \ref{WorkParams} we have left open the possibility that 
changes in ${{X_m}}$ may alter 
probabilities directly, e.g. by making new volume available for free expansion 
rather than simply via their effect on the state parameters ${{e_{{ri}}}}$.
This allows us to mathematically incorporate ``irreversible'' 
changes in entropy by averaging this term over all work parameters
and all constraints...
\begin{equation}
{\frac{{{{\delta S}}_{{irr}}}}{k}\equiv \sum _{r=1}^{R}{{\lambda }_r}\sum _{m=1}^{M}{{{\delta X}}_m}\sum _{i=1}^{{\Omega }}{e_{{ri}}}{{\left(\frac{{{{\partial p}}_i}}{{{{\partial X}}_m}}\right)}}.}
\end{equation}

If we further define ``heat increments'' ${{{{\delta Q}}_r}}$ of the rth type as... 
\begin{equation}
{{{{\delta Q}}_r}\equiv \sum _{i=1}^{\Omega }{e_{{ri}}}\sum _{u=1}^{R}{{{\delta \lambda }}_r}{{\left( \frac{{{{\partial p}}_i}}{{{{\partial \lambda}}_u}}\right)}}, \forall r\in \{1,R\},}
\label{HeatIncr}
\end{equation}
a couple of familiar differential relationships follow...
\begin{equation}
{{{{\delta Q}}_r}-{{{\delta W}}_r}=\sum_{i=1}^{\Omega }\left({e_{{ri}}}{{{\delta p}}_i}+{p_i}{{{\delta e}}_{{ri}}}\right)={{{\delta E}}_r}, \forall r\in \{1,R\},}
\label{MELaw1}
\end{equation}
and 
\begin{equation}
{\sum _{r=1}^{R}{{\lambda }_r}{{{\delta Q}}_r}+\frac{{{{\delta S}}_{{irr}}}}{k}=\sum _{i=1}^{\Omega }{{{\delta p}}_i}\sum _{r=1}^{R}{{\lambda }_r}{e_{{ri}}}=\frac{{\delta S}}{k}}.
\label{MELaw2}
\end{equation}
These are more general forms of the open system $1^{st}$ and $2^{nd}$ Laws 
(equations \ref{Law1open} and \ref{Law2open}), based purely on statistical inference 
from a description of allowed states.  The familiar physics only arrives, e.g. 
for the canonical ensemble case when $R=1$, if we further postulate that ${{E_1}}$ 
represents a conserved quantity in transfers between systems, and that 
${{{{\delta S}}_{{irr}}}}\geq 0$.
 
\subsubsection{Symmetry between ensembles}

Different thermodynamic ``ensembles'' often relegate a work parameter $X_m$ to 
the status of a constraint $E_r$ by expansion of the state sum to include all 
possible values for the work parameter.  The classic example is the pressure 
ensemble mentioned above, in which the traditional work parameter volume $V$ is 
introduced as a constraint enabling, for example, a study of volume fluctuations.
The symmetry of the equations with respect to these quantities might be better 
seen if we define M ``work multipliers'' ${{J_m}}$, analogous to the R 
``heat multipliers'' ${\lambda}_r$, as averages over all constraints 
$E_r$ of the rate at which the $e_{ri}$ depend on the work parameters 
to which they correspond, i.e.
\begin{equation}
{{J_m}\equiv \sum _{r=1}^{R}{{\lambda }_r}\sum _{i=1}^{\Omega }{p_i}{{\left( -\frac{{{{\partial e}}_{{ri}}}}{{{{\partial X}}_m}}\right)}}, \forall m\in \{1,M\}}
\label{WorkMult}
\end{equation}
Then we can also write...
\begin{equation}
{\sum _{m=1}^{M}{J_m}{{{\delta X}}_m}=\frac{{\delta S}}{k}-\sum _{r=1}^{R}{{\lambda }_r}{{{\delta E}}_r}=-\frac{\delta A}{k}+\sum _{r=1}^{R}{E_r}{{{\delta \lambda}}_r}.}
\label{Partials}
\end{equation}
...yielding a harvest of partial derivative relationships for entropy and availability.  

These include the connection between multipliers (like reciprocal temperature) 
and entropy derivatives: 
\begin{equation}
J_m={\left( \frac{\partial S/k}{\partial X_m}\right)}_{X_{s\neq m},E_r}\text{, } 
\lambda_r={\left( \frac{\partial S/k}{\partial E_r}\right)}_{E_{s\neq r},X_m}
\label{DegsFreedom}
\end{equation}
which for example allow one to show that integral heat capacities like 
$\frac{U}{kT}$ are logarithmic entropy derivatives (or ``multiplicity 
exponents'') of the form $U \frac{\partial S}{\partial U}$ taken under 
no-work conditions.  Equation \ref{MELaw2} allows one to show more 
generally that integral heat capacities like $\frac{H}{kT}$ are also 
multiplicity exponents of the $E_r$, and that differential heat capacities 
taken with $\delta S_{irr}=0$, e.g. of the form $\frac{\delta H}{\delta{kT}}$, 
are in a complementary way multiplicity exponents of the $\lambda_r$, e.g. 
of the form $T \frac{\partial S}{\partial T}$.

The ``availability slope'' partials relate fluctuating 
parameters $J_m$ and $E_r$ to control parameter partials 
taken under ensemble conditions, i.e. 
\begin{equation}
J_m=-{\left( \frac{\partial A/k}{\partial X_m}\right)}_{X_{s\neq m},\lambda_r}\text{, } 
E_r={\left( \frac{\partial A/k}{\partial \lambda_r}\right)_{\lambda_{s\neq r},X_m}}.
\label{AvailabilitySlopes}
\end{equation}
These are the starting point for our assertion, in the abstract, about 
the general usefulness of uncertainty slopes in problems of 
statistical inference.  The next section takes the assertion a step 
further, by providing insight into fluctuations and correlations.

\subsubsection{Fluctuations and reciprocity}

Again following Jaynes and taking partials under ensemble constraints, given 
\begin{equation}
\frac{{{{\partial p}}_i}}{{{{\partial \lambda }}_r}}= p_i ( E_r - e_{ri}) \text{, }\forall i\in \{1,\Omega\} \text{ \& }\forall r\in \{1,R\},
\end{equation} 
and
\begin{equation}
{\delta p}_i=\sum _{r=1}^{R}{\delta \lambda}_r \frac{\partial p_i}{\partial \lambda_r}+\sum _{m=1}^{M} \delta X_m \frac{\partial p_i}{dX_m}, \forall i\in \{1,\Omega\},
\end{equation}
one can show quite generally that the covariance between parameters $E_r$ and $E_s$ 
(namely $\langle {e_r}{e_s}\rangle -\langle {e_r}\rangle \langle {e_s}\rangle$) is minus the partial of $E_r$ with respect to $\lambda_r$, and hence that
\begin{equation}
{\sigma^2}_{E_r E_s}=
-\frac{\partial E_r}{\partial \lambda_s}=
-\frac{\partial E_s}{\partial \lambda_r}=
-\frac{\partial^2 A/k}{{\partial \lambda_r}{\partial \lambda_s}},\forall r,s\in
\{1,R\}.
\label{Covariance}
\end{equation}
The latter equality gives rise to the Onsager reciprocity relations 
of non-equilibrium thermodynamics.  

For the special case when $r=s$, the above
expression also sets the variance (standard deviation squared) 
of $r$ to $\sigma_{E_r}^2=-\frac{\partial E_r}{\partial \lambda_r}$.
Since the left hand side of this equation seems likely positive, 
the equation says that temperature, for example, is likely to increase with
increasing energy.  This turns out to be true even for systems like spin 
systems which exhibit negative absolute temperatures, provided we recognize
that negative absolute temperatures are in fact higher than positive absolute 
temperatures i.e. that the relative size of temperatures must be determined
from their reciprocal ($1/kT$) ordering.  Since this quantity is also 
proportional to heat capacity, equation \ref{Covariance} also 
says that when heat capacity is singular (e.g. durng a first order phase 
change), the fluctuation spectrum will experience a spike as well.

Although we only conjecture based on symmetry here, similar 
relations may also obtain for the work multipliers, e.g. 
\begin{equation}
{\sigma^2}_{J_m J_n}=
-\frac{\partial J_n}{\partial X_m}=
-\frac{\partial J_m}{\partial X_n}=
\frac{\partial^2 A/k}{{\partial X_m}{\partial X_n}},\forall m,n\in
\{1,M\}.
\label{JCovariance}
\end{equation}
as well as for hybrid multiplier covariances.  

\subsubsection{Net surprisal \& availability}
\label{subsec:MaxEnt6}

Changes in availability under ensemble constraints, as 
in the derivatives above, can also be seen as whole
system changes in uncertainty relative to a reference 
state \cite{Evans69, Tribus71}, i.e. as changes in 
net surprisal.  Here we define net surprisal as
\begin{equation}
I_{net} \equiv - k \sum _{i=1}^{\Omega }{p_i}\ln (\frac{p_{oi}}{p_i}) \geq 0\text{,}
\label{NetSuprisal}
\end{equation}
where the $p_{oi}$ are state probabilities based only 
on ambient state information, while the $p_i$ take into 
account all that is known.  The inequality follows 
simply since each set of probabilities contains 
only positive values that add to one.  It then follows 
from equations \ref{Solution}, \ref{PartitionFn} and \ref{MaxEnt} that 
\begin{equation}
\frac{I_{net}}{k} = -(\frac{S}{k} - \frac{S_o}{k}) + \sum_{r=1}^{R} (E_r - E_{ro}) \lambda_{ro} \text{.}
\label{NetSuprisal2}
\end{equation}
provided our 
system's deviation from the reference state (here no longer 
infinitesimal but finite) does not involve changes in the 
work parameters $X_m$, since this would constitute a 
change in the problem (e.g. the energy level structure) 
being considered.  If the $E_r$ are conserved in transfer 
between systems, net surprisal is simply the entropy 
increase of {\em system plus environment} on equilibration to 
ambient, with the surprisal value of excess $E_r$ simply
the ambient uncertainty slope $\lambda_{ro}$ (e.g. 
$1/T_o$ for energy $U$).  Using equation \ref{AvailBits} 
we then get
\begin{equation}
\frac{I_{net}}{k} = (\frac{A}{k} - \frac{A_o}{k}) - \sum_{r=1}^{R} (\lambda_r - \lambda_{ro}) E_r \text{.}
\label{NetSuprisal3}
\end{equation} 
Thus near ambient conditions, derviatives of availability under ensemble 
conditions are also derivatives of net surprisal.

For example, systems in thermal contact with an 
ambient temperature bath may be treated as canonical 
ensemble systems with constrained average energy.  
Thus a temperature deviation from ambient $T_o$ for a 
monatomic ideal gas gives for that system 
$\frac{I_{net}}{k} = \frac{3}{2} N \Theta [\frac{T}{T_o}]$ where 
$\Theta [x] \equiv x - 1 -\ln x \geq 0$.  If that system is also 
in contact with an ambient pressure bath (i.e. able to 
randomly share volume and energy), volume deviations add 
$N \Theta [\frac{V}{V_o}]$ to the foregoing.  For grand canonical 
systems whose molecule types might change (e.g. 
via chemical reaction), one instead adds $N_j \Theta [\frac{N_{jo}}{N_j}]$  
for each molecule type $j$ whose concentration varies 
from ambient. 

Not only is the concept of net surprisal simply represented in 
context of the general maxent calculation, it also offers 
simplifying insight into thermodynamic processes.  For example, 
an {\em at first glance counter-intuitive} problem offered to 
intro physics students at the University of Illinois asks 
how cold the room must be for an otherwise unpowered device to
take boiling water in at the top, only to return it as ice water 
with a bit of ice therein at the bottom.  Since the $2^{nd}$ law 
allows conversion of one form of net surprisal reversibly 
into another (famously without a clue how to do it in practice), 
one can use the fact that the function $\Theta $ above works for 
``quadratic systems'' in general to set 
$C_v \Theta [\frac{T_{hot}}{T_{room}}] = C_v \Theta [\frac{T_{ice}}{T_{room}}]$ 
and solve for $T_{room}$.  Thus with net surprisal in hand the 
problem becomes both conceptually, and analytically, simple.

Of course, the net surprisal measure is not only relevant to 
inference about systems for which physically conserved energy 
is of interest (i.e. thermodynamic systems).  In fact, one 
might conjecture that it meets the requirements for an information 
measure proposed by Gell-Mann and Lloyd \cite{GellMann96}, and 
that it includes the Shannon information measure discussed there as a 
special case.  By way of a specific application, net surprisal's 
usefulness for quantifying the amount of information students bring 
to an exam is illustrated in Appendix \ref{app:MultipleChoice}.  Thus 
armed with statistical inference tools that underpin 
traditional thermodynamic applications, but which require no physical 
assumptions {\em a priori} short of a state inventory, we now take a look 
at some of the more complex system areas where applications (already 
underway in many fields) will likely continue to develop.

\section{Steady-state engines}
\label{sec:Engines}

  Begin by considering within our larger isolated system the possibility of 
``steady-state engines'', i.e. devices which operate in some fashion on 
their surroundings while remaining (to first order) the same themselves 
before and after.  If we refer to $U_i$ and $S_i$ respectively as the 
steady state energy and entropy of engine $i$, then the change with time 
of these values will be (by definition) negligible.  Hence such steady 
state engines contribute little or nothing to time variations in total 
system energy and entropy.  Hence the 1st and 2nd Laws applied to 
engines plus environment means that the same equations apply to the 
energy $U$ and entropy $S$ {\em external} to such steady-state systems 
alone.

  Since energy and work can be exchanged in both ways between our 
engines and their environment, it is convenient to write:
\begin{equation}
dU = (\delta Q_{out} + \delta W_{out}) - (\delta Q_{in} + \delta W_{in}) = 0\text{, and}
\label{SSLaw1}
\end{equation}
\begin{equation}
dS = \frac{\delta Q_{out}}{T_{out}} - \frac{\delta Q_{in}}{T_{in}} = \delta S_{irr} \geq 0\text{.}
\label{SSLaw2}
\end{equation}
Here the terms with the subscript ``in'' represent flows of energy into our 
steady state engines, while the terms with the subscript ``out'' represent flows of 
energy out from those same engines.  These equations open the door for students 
to a wide variety of ``thermodynamic possibility'' calculations.  Heat and
information engines will be our focus here.

\subsection{Heat engines \& biomass creation}
\label{subsec:HeatEngines}

\begin{figure}
\includegraphics{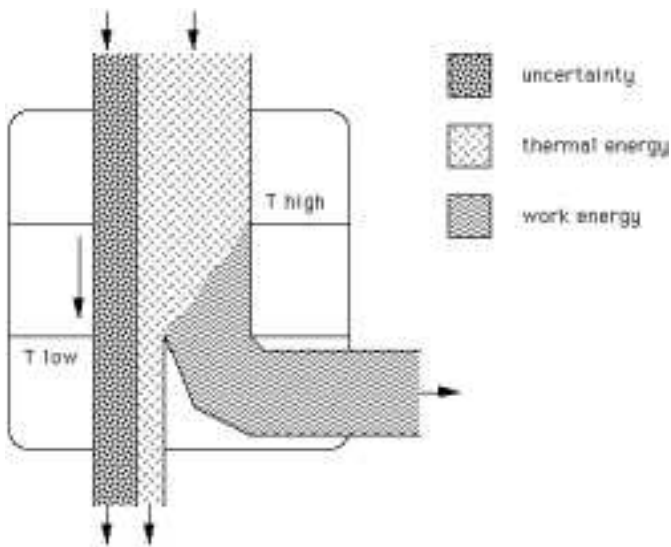}%
\caption{Schematic for a heat engine which brings in heat energy 
at high temperature and returns heat at low temperature along with 
energy available for work.  One example, applicable to 
automobile engines, is the lifting of a weighted piston with hot 
gas which cools on expansion.  Another is the storage of available 
work in plant biomass via photosynthesis of radiation from the sun.}
\label{Fig1}
\end{figure}

Heat engines as in Fig. \ref{Fig1} are generally defined as steady-state systems which 
take in heat from a high temperature (e.g. a combustion) reservoir, and return 
that energy as heat and ``ordered energy'' to a lower temperature (e.g. ambient) 
reservoir.  Car and steam engines fall into this category, if we allow burning 
fuel to be considered their source of high temperature heat.  

The equations above also work with forms of plant life which take in sunlight (high 
temperature heat) and store chemical energy (i.e. work) in plant biomass (e.g. 
in cellulose, carbohydrates, proteins, and fats).  In this case, $P dV$ work may 
be ignored, and $\delta W_{out} - \delta W_{in}$ becomes the change in chemical 
potential times the number of molecules whose state is changed by solar irradiation.
Ecologists refer to organisms that do this as {\em autotrophs} (``self-nourishers'') 
or primary producers \cite{Odum71}. 

The exhaust (i.e. low temperature) reservoir for most heat engines is 
the ambient environment.  Refrigerators and electric heat pumps are by 
comparison heat engines run in reverse, i.e. they take in work and 
heat from a low temperature reservoir, exhausting it to a warmer ambient. 
All of the exhausted heat is eventually radiated at around 300K back 
into space by the earth, letting us see the earth itself as a 
steady-state heat engine as well.

Solving equations \ref{SSLaw1} and \ref{SSLaw2}, and assuming that $T_{in}$ is positive, 
we get the familiar upper limit on energy available for work that a Carnot engine 
(i.e. an engine whose heat flows out of and into a pair of fixed-temperature reservoirs) 
can produce:
\begin{equation}
\delta W_{out} \leq \delta Q_{in} \left( 1 - \frac{T_{out}}{T_{in}} \right) \text{.} 
\label{Carnot}
\end{equation}
Most real heat engines have efficiencies (conversion fractions) which are beneath 
this because of irreversibilities during operation.

\subsection{Information engines \& us}
\label{subsec:InformationEngines}

The concepts of thermodynamics have been traditionally honed in 
systems near or approaching equilibrium, and the entropy of 
homogeneous systems at equilibrium is an extensive quantity like 
energy or volume or number of particles.  However, the 
maximum entropy best guess machine is much less restrictive 
about the kinds of system to which it applies.  In particular, 
uncertainty about the state of a system in general depends not 
only on what we know about each component of a system, but 
what we know about the relationship between components.  

For example, if 10 white and 10 black marbles are distributed between 
two drawers A and B, then one has $S=20 k \ln 2$, or 20 bits, of 
uncertainty about the drawer assignment of these marbles (i.e. one 
true-false question's worth, or bit, of uncertainty per marble).  
However, if one is given as true the statement that 
``marbles in any given drawer are all the same color'', the 
uncertainty about the drawer assignment of marbles is 
reduced to $k \ln 2$ or one bit of uncertainty.  Even though 
a bit (literally) of uncertainty remains about the drawer assignment 
for each individual marble, as before, the total uncertainty 
has now been decreased by the mutual information in that 
statement, or
\begin{equation}
M = \sum_{i=1}^{N_{ss}} S_i - S_{total} \text{.}
\label{mutual_inf}
\end{equation}
In our example, there are $N_{ss}=20$ subsystems, and this equation 
shows that $M = 20 - 1 = 19$ bits of mutual information are 
contained in the statement quoted above!

Mutual information (e.g. that two spins are correlated, or 
that two gases have not been well mixed) plays a well-known 
role in physical systems as well \cite{Brilloun62, Lloyd89b, Scully03}, 
with recent focus in particular on it's impact 
in nucleic acid replication \cite{Schneider91b, Schneider00} 
and in quantum computing \cite{Wootters82, Gottesman00}.  For example, 
Grosse et al  \cite{Grosse00} use intra-molecule mutual information to 
distinguish coding and non-coding DNA, instead of autocorrelation 
functions, because the former does not require mapping symbols to numbers, 
and because it is sensitive to non-linear and linear dependences.  
Although constraints of this sort may be incorporated into the maxent 
formalism (cf. Appendix \ref{app:MutualInformation}), we take the 
possibility of such correlations into account here by simply modifying 
equation \ref{Law2open} to read 
\begin{equation}
\delta S = \frac{\delta Q_{in}}{T} - \delta M_{internal} + \delta{S_{irr}} \text{.}
\label{Law2i}
\end{equation}
This makes the $2^{nd}$ Law relevant to engines whose primary function 
concerns tasks not explicitly involving changes in energy, 
such as the job of putting ``the kids' socks in one pile and 
the parents' socks in another'', or the challenge of reversible 
computing.  When $\delta M_{internal}$ changes
are important, however, note that entropy cannot be considered 
an extensive quantity like U, V, and N, since the total 
uncertainty S about the state of a system may be less than 
the sum of the uncertainties about the state of its constituent 
parts.

This strategy reflects recent thinking about the energy cost 
of information in generalizing the Maxwell's demon problem \cite{Bennett87}. 
Zurek \cite{Zurek89} among others suggests that the only {\em requisite} 
cost of recording information about other components in a system 
is the cost of preparing the blank sheet (or resetting the measuring 
apparatus} prior to recording with it.  Moreover, the minimum 
thermodynamic cost, in energy per unit of correlation information, 
is simply the ambient temperature $T$.  

\begin{figure}
\includegraphics{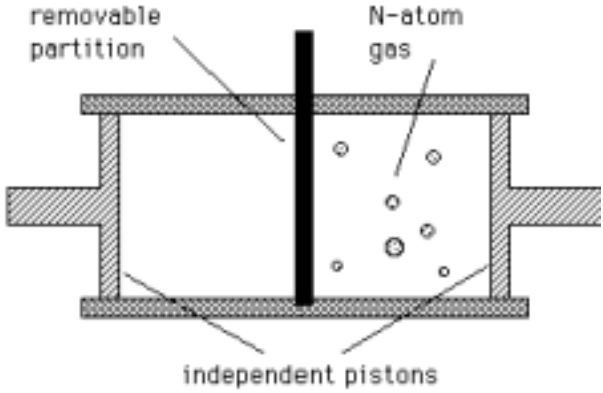}%
\caption{A symmetric bi-partitioned cell for the 
isothermal compression of an N-atom ideal gas into either 
its left or right half, perhaps first discussed by Szilard, 
which serves as a physical system about whose state 
mutual information is available for a well-defined price 
in free energy.  If one is further provided with some 
mechanism (e.g. spectroscopic) for reading its state, it 
may also serve as a mechanically operated single-bit 
memory.  The ``setting'' process involves removing the 
barrier between compartments, using the piston on one side 
to relocate all atoms into the opposite half and then 
returning the barrier before returning the piston to its 
original position.  The required work is $W_{in} = N k T \ln[2]$. 
Its reset status may be defined as true if we know that the 
atoms are located in the right half of the container, and 
false if we don't know this to be the case.}
\label{Fig2}
\end{figure}

A classic example \cite{Penrose70} of this is the isothermal compressor 
for an N-atom gas \ref{Fig2}, taken for the case when N=1.  The system 
requires thermalization of no less than $k T \ln 2$ of work, in return 
for a single bit of correlation information concerning the location of 
the atom.  That correlation information in turn can be used 
subsequently to perform {\em with arbitrarily small energy cost} the 
same task of locating the atom on a desired side, by simply rotating 
the cylinder by 180 degrees as needed!

The $\Delta M$ term also lets us see the {\em isolated} system 
Second Law (Eqn \ref{Law2}) in a new light.  Begin with a system A 
with $\Omega$ total accessible states, so that uncertainty about A is at most 
$S[A] = k \ln \Omega $.  Then consider an observer B, with sufficient added information 
about A to limit the number of accessible states to $\Gamma  < \Omega $.  
Observer B therefore has conditional uncertainty about A (see Appendix 
\ref{app:MutualInformation}) of $S_B[A] = k \ln \Gamma $.  
What we can learn about A by knowing B is then the mutual information 
$M[A,B]=S[A]+S[B]-S[AB]=S[A]-S_B[A]=k \ln \frac{\Omega}{\Gamma}$.  
If the basic structure of system A from which $\Omega$ was calculated 
remains intact, then the isolated system $2^{nd}$ Law assertion that 
observer uncertainty about isolated A can only stay constant or increase 
(i.e. that $dS_B[A]/dt \geq 0$) implies also that the mutual information 
between two isolated systems (here $M[A,B]$) can only decrease.

\begin{figure}
\includegraphics{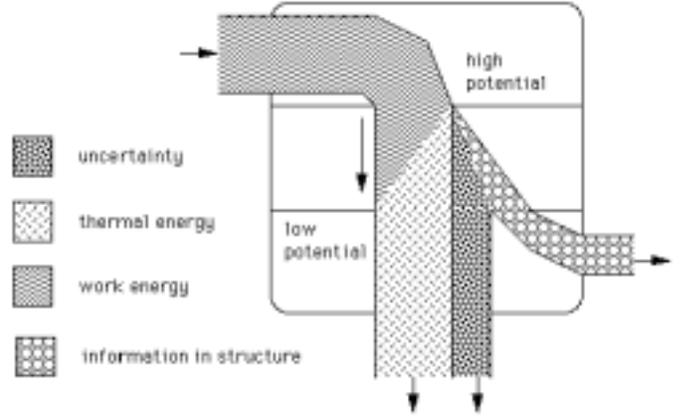}%
\caption{Schematic for an information engine which in the process 
of thermalizing energy available for work creates mutual information 
(also called correlation information, information in structure, or 
negentropy \cite{Brilloun62}).  One example is the isothermal compression 
of an N-atom 
gas into the left half of a compartmented container with a vacuum 
pump, while thermalizing at least $N k T \ln[2]$ of work energy and 
creating some external record of the occurrence.  Another example 
might be the resetting (erasure) of a used sheet of paper or a 
molecular template, as a first step in the encoding or replication 
of a message.}
\label{Fig3}
\end{figure}

Now we consider steady state engines whose function is to produce  
mutual information or correlations between two systems, as 
in Fig \ref{Fig3}.  These correlations might, for example, be 
marble collections sorted by color, a faithful copy of a strand of 
DNA, or dots on a sky map corresponding to the position of stars in 
the night sky.  Our first and second law engine equations (from 
\ref{SSLaw1} and \ref{SSLaw2} with mutual information), become
\begin{equation}
dU = (\delta Q_{out}) - (\delta W_{in}) = 0\text{, and}
\label{IELaw1}
\end{equation}
\begin{equation}
\delta S = \frac{\delta Q_{out}}{T_{out}} - \delta M_{external} = \delta S_{irr} \geq 0\text{.}
\label{IELaw2}
\end{equation}
Eliminating $Q_{out}$ from these two equations yields 
\begin{equation}
\delta M_{external} \leq \frac{\delta W_{in}}{T}\text{.}
\label{IEresult}
\end{equation}
This means that information engines can produce no more mutual 
information than their energy consumption, divided by 
their ambient operating temperature.  In binary information units, 
this amounts to producing about 55 bits of information per eV of 
thermalized work at room temperature, and around 60 bits per eV 
of energy if operating near the freezing point of water.

Cameras, tape recorders, and copying machines may be considered 
such information engines, as are forms of life which take in 
chemical energy available for work from plant biomass, and thermalize 
that energy at ambient temperature while creating correlations 
between objects in their environment and their survival needs, 
and in the form of persistent DNA sequences, behavior redirections, 
songs, rituals, books, and sets of ideas).  Living organisms which 
do not qualify as heat engines, but which fit this description, 
are known by ecologists as heterotrophs \cite{Odum71}, a category 
which includes most non-photosynthetic organisms (including humans).

For a human being consuming $1.3 \times 10^7$ joules (around 3000 kcal 
or 7 twinkies) per day, and viewed as an information engine, this 
implies an upper limit on production of $4 \times 10^{27}$ bits of
mutual information in our environment per day.  (Note:  This includes 
non-coded correlations, like laundry which has been sorted into 
piles, as well as coded correlations such as a map of the night 
sky as it looked at 11pm from your backyard.) 
Much like conservation of mechanical energy limits on speed in 
roller coaster rides, and Gauss's law limits on net charge within a 
volume based on field measurements at the boundaries, this assertion 
may well stand quite independent of the detailed biochemistry going 
on inside.  Alas some of us, in practice, have trouble putting a one 
page report of new observations in a file per week!  Although individual 
metazoans in fact bolster the correlation information in their environment 
on many levels (see the next section), even when unassisted by other 
sources of energy available for work, it is likely that the above 
inequality is not a major bottleneck for most.

\section{Excitations and codes}
\label{sec:CodesExcitations}

\begin{figure}
\includegraphics{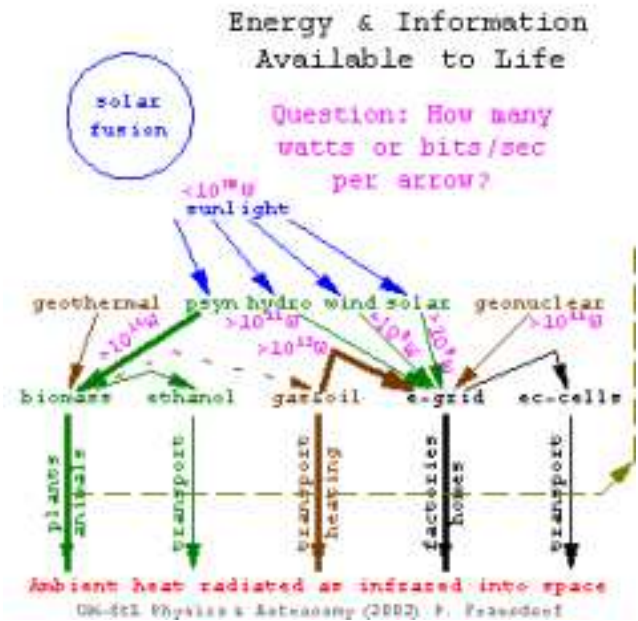}%
\caption{Life's Energy Flow: The top half represents some of 
the primary processes involved with energy flow, while the 
bottom half illustrates significant physical repositories for 
life's energies, and paths for conversion of energy from one
form to another.}
\label{EnergyFlow}
\end{figure}

Figure \ref{EnergyFlow} illustrates the flow of energy available 
to life on earth, much of which began as high temperature 
radiative heat given off by the sun, subsequently converted 
to chemical potential energy (heat engine style) in the form of 
plant biomass \cite{Odum71}.  Much work might still be done to quantify 
these flows \cite{Vitousek97}, since the flow rate through 
biomass is seldom even considered outside of classes in 
ecology.  For example, many introductory physics texts, and even the 
world almanac, ignore its size entirely.  Hence student 
projects on {\em the size} of these flows, at various times and 
places, might be interesting and enjoyable.  Likewise for 
projects which examine the involvement of various consumables 
and activities in the depicted streams e.g. the availability 
cost of a hot dog, or an aluminum can.

Some of the ''ordered energy'' outputs from the heat engines 
described in Figure \ref{EnergyFlow} are eventually 
thermalized (e.g. in forest fires or the burning of fossil 
fuels), but not all of it is irreversibly thermalized.  
In other words, some of the free energy made available by 
plants is converted to non-energy related correlations 
between organism and environment, and some is converted to 
to internal correlations within living things.  

Organism/environment correlations include, for example, 
cell membranes that separate the contents of one-celled organisms 
from the fluids surrounding.  They differ, depending on the 
nature of the ambient to which a given organism is adapted.  
Similarly, the woody trunks of trees don't merely store 
chemical energy for later combustion, but instead point in 
a direction which allows subsequent leaf growth to have better 
access to the light of the sun.  

\begin{figure}
\includegraphics{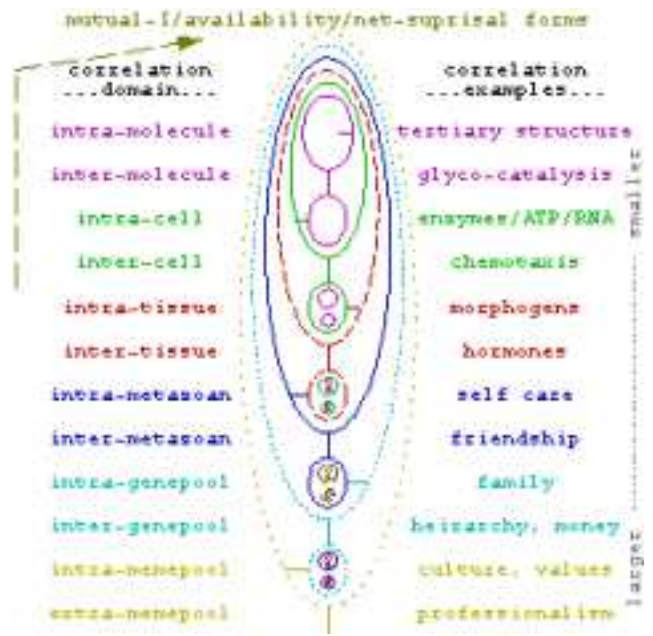}%
\caption{Life's Stores of Availability: Horizontal bars 
represent inward-looking (intra or ``yin'') correlations, while 
vertical bars represent outward-looking (cross-boundary or 
``yang'') correlations.  This breakdown seems to work 
reasonably well to categorize by domain both the types of 
correlations that exist, and the kinds of ideas (i.e. memetic 
replicators) used to help maintain them.}
\label{SuprisalMap}
\end{figure}

Correlations internal to organisms include catalysts (often 
amino acid enzymes) which guide the spending of the cell's 
energy coin \cite{Hoagland98} (adenosine triphospate molecules) 
not only toward 
nourishment and other external goals, but toward it's use 
in the process of cell replication.  The enzymes themselves 
are typically constructed from amino acid sequences which 
fold in solution into secondary and delicate tertiary 
structures which are crucial to catalyst structure and 
function.

In fact, information on these correlations within catalyst 
molecules, resulting in part also in correlations between 
organisms and their environment, apparently proved so important 
that a {\em digital} means (nucleic acid codes) to store mutual 
information on these correlations, still in widespread use 
today, was developed several billion years ago \cite{Margulis86, Ward00}.  
Note that the word digital here refers to ways to store mutual 
information in which bit-wise fidelity of the replication process 
can be checked after the fact.  This is distinct from {\em analog}
forms of recording, like the storage of images on film, 
where accuracy on the microscopic scale is lost statistically 
in the grain structure of the film, as one moves to 
increasingly smaller size scales.  

This ancient invention of digital recording more or less
formalized a now long-standing symbiosis between {\em steady-state 
excitations} (in particular organisms which operate in-part 
by reversibly thermalizing an inward-flowing stream of energy in the 
form of available work) and {\em replicable codes}.  This excitation-code 
symbiosis, of course, involves mutual information managed (stored, 
replicated, and applied to enzyme manufacture) by biological cells \cite{Hoagland98}. 
Now memetic replicators \cite{Dawkins89, Blackmore00}, i.e. ideas 
which began as sharable patterns stored in the neural nets of 
animals \cite{Dennett92, Schaik03}, are in the process of going digital 
\cite{Mcluhan62, Harnad91}, thus adding a second level to life's 
symbiosis with replicators.  The unconscious struggle for 
hegemony over organisms, between these two replicator families, 
might in a way be seen as a battle between ``sword'' and ``pen'' 
in which (strangely enough) organisms are the spoils of war.
At the very least, it seems likely that under some conditions
the interests of organisms, and the interests of codes, don't 
commute.

Naturally our ``organism-centric'' vantage point prompts us to 
miss, at first glance, the way that organisms serve codes in a 
given process \cite{Dawkins89}.  We might sometimes even miss the 
distinction between ``our ideas about the world'' (those replicable codes) 
and ``the world itself'' (a complex excitation with deep internal 
correlations) \cite{Hofstadter85} as though we are in danger of 
``knowing everything'' with a completed map of the universe in our 
minds \cite{Horgan96}.  A closer look at nature, however, reveals 
that true cloning of internally-correlated excitations (e.g. like 
qubits \cite{Wootters82}) may be impossible in principle as well as 
in practice.
 
A natural way to ``illustrate and inventory'' the standing crop of 
correlations associated with life, while recognizing boundaries 
between replicator-pools as well as simpler physical boundaries (like 
cell walls and individual spaces), is illustrated in Figure 
\ref{SuprisalMap}.  Again, students might find it 
enjoyable to think about ways to inventory this standing 
crop of correlations, at various places and times.  Although 
in principle each bar in the figure could be quantified in 
``bits of standing availability'', neither the means nor the 
motivation for doing this objectively are clear at this point.  
However, just picturing qualitatively the state of these 
correlations and the boundaries with which they associate, 
as physical elements in the world around, might be worthwhile
(cf. Fig \ref{BoundaryMap}).

\begin{figure}
\includegraphics{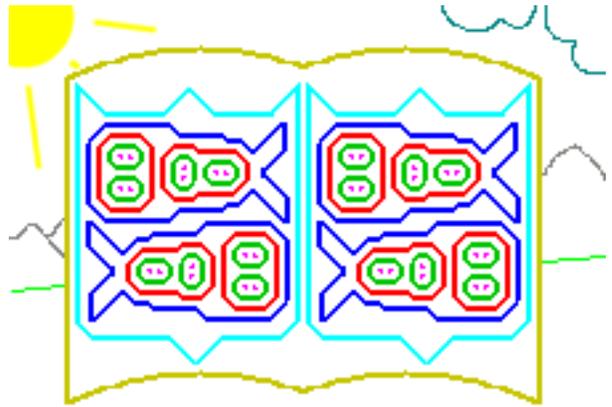}%
\caption{An expanded schematic of some boundaries alone, 
drawn from Fig \ref{SuprisalMap}.  This suggests for 
example that the idea sets for interaction between 
cultures (i.e. when one has to take into account more 
than one book in the figure, or meme-pool) may be quite 
different than those for interactions between heirarchies 
in a given culture.  Guidelines for professionalism in 
the workplace, and political correctness, as irrelevant 
as they may seem in the context of one culture, might be 
evidence of behavior correlations developing along those 
lines.}
\label{BoundaryMap}
\end{figure}

\section{The natural history of invention}
\label{sec:Invention}

Histories of emergent phenomena, like 
Marshall McLuhan's ``Gutenberg Galaxy'' \cite{Mcluhan62}, 
Konrad Lorenz' ``On Aggression'' \cite{Lorenz66}, 
Margulis \& Sagan's ``Microcosmos'' \cite{Margulis86}, 
Jared Diamond's ``Guns Germs and Steel'' \cite{Diamond97}, 
David Attenborough's Special on Birds \cite{Attenborough97}, and 
Ward \& Brownlee's ``Rare Earth'' \cite{Ward00} (in broad strokes at 
least) simplify when outlined in terms of the two manifestations 
of generalized availability depicted in Figures \ref{EnergyFlow} 
and \ref{SuprisalMap}: (i) ``ordered'' or free energy, and (ii) mutual information, 
respectively.  These two themes repeatedly intertwine in a non-repeating drama that 
involves partnership between replicable codes (which incidentally include the above 
two concepts), and what physicists might call steady-state excitations busily 
converting energy and information from one form to another.  This history shows 
potential for providing a neutral perspective, grounded in established physical 
and logical principles, on numerous important and sometimes contentious issues.  
By providing context both for such issues and our reactions to them, it might 
catalyze constructive dialog. It also suggests elements of a natural history, 
informed to interdisciplinary connections emergent only in the last century.  

A larger ``timeline of concept-relevance'' for ideas might thus, for example, 
begin with the elemental concepts of:

\begin{itemize}
\item dimension (1D, 2D, 3D, 3+1D, n+1D, etc.),
\item metric (e.g. Pythagoras' space \& Minkowski's space-time);
\end{itemize}

followed by the emergence in our world of manifestations now represented 
by basic physical concepts like:

\begin{itemize}
\item motion, momentum-energy, mass \& gravity
\item other interactions like charge \& electromagnetic force,
\item particles, waves, atoms \& their associated chemistry,
\item heat, an early application of gambling with uncertainty,
\item {\em available work} and {\em mutual information} in physical systems.
\end{itemize}

Discoveries on earth then lead to the following inventions by one-celled organisms:

\begin{itemize}
\item steady-state {\em thermal-to-chemical} (i.e. photosynthetic) and {\em chemical 
to kinetic} (i.e. motion) energy conversion,
\item {\em energy storage} in combustable sugars, and mechanisms for withdrawal (via enzymes) into more easily and universally spendable ATP molecules.
\item {\em digital information partnering} (trading work for recorded correlations) 
with highly replicable amino and nucleic acid codes, and thus in a sense the practice 
(but not yet the idea) of genetic engineering;
\item {\em intracellular structures} like cell membranes \& organelles (symbiotic) and viruses (parasitic),
\item chemical and tactile {\em messaging};
\item tools like thermal or chemical gradients for locating energy sources, \& contact forces for motility;
\end{itemize}

subsequent inventions by multi-celled plants of:

\begin{itemize}
\item {\em intercellular correlations} like eukaryotic cells, sexual reproduction \& microbe-assisted digestion (symbiotic), or bacterial infections (parasitic),
\item {\em differentiated structures} like circulatory systems, leaves, stalks, roots, flowers, \& shells,
\item {\em other intra-organism correlations} such as annual/biennial reproductive cycles (symbiotic), or cancerous tissues (parasitic), and
\item {\em inter-organism correlations}, like ritualized interspecies redirection of behavior by providing animals with fruit and nectar symbiotically, or fake sex parasitically, so as to distribute seeds \& pollen,
\item {\em messaging} via hormone (intra-organism) \& exterior design;
\item tools like gravity and wind as aids to reproduction;
\end{itemize}

the invention by animals of:

\begin{itemize}
\item {\em information partnering} with neural net patterns via sense-mediated action, of limited replicability perhaps greatest in ritualized songs, discovery dances, \& warning vocalizations (especially for birds, bees, and mammals),
\item {\em intra-organism structures} like vertebrae, muscles, brains, eyes with lenses, gills, lungs, hearts, legs \& wings,
\item intra-species aggression and it's ritualized redirections \cite{Lorenz66}, including greeting ceremonies, pair bonds \& laughter,
\item {\em family systems} serving inward-looking perspectives with respect to intra-specific gene-pool boundaries, and related correlations like joint-parenting \& constructive sibling interactions (e.g. play between kittens),
\item {\em political systems} serving outward-looking perspectives with respect to intra-specific gene-pool boundaries, and associated correlations such as heirarchy in a wolf pack,
\item {\em inter-organism messaging} via sound, body language, and interior design (bauer),
\item tools like webs, levers, vines, tunnel, dam, wood, \& stone;
\end{itemize}

and finally the invention in human communities of:

\begin{itemize} 
\item {\em information partnering} with highly replicable spoken languages, print, and most recently digital codes,
\item {\em available work production \& distribution} in these forms: food/drink, ritual (monetary), fossil fuel \& electrical,
\item subsystem repair/augmentation networks like medicine, dentistry, pharmacy, auto repair, \& physical therapy,
\item {\em redirective elicitors} (symbiotic \& parasitic) of innate behavior (like eating, procreation \& militant enthusiasm) include sports, ``mind'' chemicals, non-reproductive sex, self-help, psychotherapy, plus artificial colors, flavors, smells \& shapes (for food \& individuals),
\item {\em prediction activities} like meteorology, gambling, insurance, digital modeling, investing, polling \& quality control,
\item {\em evolving pair, family, and heirarchy paradigms} with roots in phylogenetic \& memetic tradition, like votes, jury, public corporations, church/state separation, free press, merit/goal-based management, human rights, 
\item {\em belief systems} serving inward-looking perspectives with respect to meme-pool boundaries, and related correlations that include religions (symbiotic) \& religious colonialism, plus ethnic, cultural, \& artistic identities,
\item {\em knowledge systems} serving outward-looking perspectives with respect to meme-pool boundaries, and related correlations that include professions, ``political correctness'', secular colonialism \& peer review,
\item {\em inter-organism messaging} via music, art, writing, printing, teletype, radio, phonograph, telephone, photography, cinema, fingerprint, television, xerography, magnetic tape, bar code, optical disk, pager, sky phone \& internet,
\item tools like clothes, fire, oven, wheel, ramp, plow, weapons, skyscraper, bike, road, steam/gasoline/electric motors, car, bridge, train, boat, aircraft, match, washing machine, vacuum cleaner, flush toilet, dishwasher, grinders, mirror, glass lens, camera, camcorder, hologram, clock, artificial light, metals, ceramics, concrete, canning, polymers, gear, cam, lock, spring, rope, pulley, block \& tackle, zipper, scissors, nail, screw, irrigation, planting \& harvesting equipment, wrench, portable drill/saw, end loader, running water, gas \& electric heater \& drier (for people, food, \& clothes), elevator, crane, battery, volt-ohm meter, laser, compass, satellite, global positioning system, gyroscope, autopilot, smoke detector, fridge \& air conditioner, X-ray \& ultrasound \& MRI imaging, tele \& micro \& endo scopes, spectrometers, semiconductors, transistors, integrated circuits, computers, fiber optics, robotics, and 
{\em the idea of} polymerase chain reaction for nucleic acid sequence replication.
\end{itemize}

Thus correlations, written in nucleic or amino acid strings, have been 
developing in symbiosis with microbes since the very early days of our planet.  
Moreover, sometime since the Cambrian bloom of metazoan body types, and 
particularly among humans in the past 10,000 years, similar correlations 
written in memetic codes have been undergoing active development.  The latter 
were of course broadcast not via the sharing of molecules, but by 
transference between neural nets through metazoan senses via performance, 
speech, script, and more recently digital means starting with the Phoenician
alphabet.

The large number of thermodynamic and information-theoretic processes in 
this list raises a question about codes that arises often today in  
context of the human genome project:  What gene is responsible for what 
features of an organism, or conversely what features of an organism does
a given nucleic acid sequence ``cause''?  The same question of course can be asked 
about memetic codes.  Has a given set of ideas been honed via experience 
with the world around us, via experience with worlds within this boundary or 
that, or does it offer little by way of connection to the world at all?  
I hope that we've shown here that in any rigorous sense such questions must 
be considered questions not about the properties of a molecule or a set of 
words in isolation, but rather questions about delocalized correlations 
between physical objects (in particular between codes or their phenotypes, 
and other parts of the world around).  
Once the context is specified (e.g. the reference state used in equation 
\ref{NetSuprisal}), objective and even quantitative assessments of these 
correlations may be possible.  

Qualitatively, for example, most might agree that the nucleic acid base triplets 
UAA, UAG, and UGA have evolved as elements of punctuation in the genetic code, 
there not to correlate with the outside world but to guide the process 
of transcription into protein, much as the period at the end of this 
sentence guides the sentence's transcription into speech.  Such punctuation 
codes are one kind of internal code, developed to guide the replication of 
codes and their reduction to practice.  Other codes have 
evolved by virtue of (i.e. their survival has been connected in a real-time 
manner to) the correlations that they affect between an organism and the 
inanimate world around.  Thus a chunk of genetic code might correlate with 
the thickness to length ratio of a plant's stem, whose optimum value may 
depend on wind velocities and topography in the world around.  Similarly, a set of 
ideas for guiding the path of a ship at sea might survive depending on its 
usefulness in helping the sailors reach their destination, before they run 
out of supplies.  

Some kinds of internal code affect (and are affected by) the way manufacturing 
is carried out within cells.  Others affect the ways cells interact with one another, 
and yet others affect the way tissues function as a unit, etc., across the levels 
illustrated in Figure \ref{SuprisalMap}.  Codes (genetic or memetic) whose survival 
is predicated primarily on correlations between or internal to lifeforms (rather 
than specifically between a lifeform and it's inanimate environment) might be 
called ``we-codes''.  Thus for example, many might agree that legal systems 
provide guidelines (in this case we-memes) for cooperation between more than 
one genetic subgroup or nuclear family.  Clarifying our ideas about the ways that 
segments of code participate in correlations between the organisms they guide, 
and other parts of the world, is even more important now that genetic codes are 
being transribed by humans into memetic form.

Examining any given correlation from this list quantitatively 
(cf. Appendix \ref{app:MutualInformation}) may or may not be 
meaningful.  However, the list does make it easy to see 
why thermodynamic metaphors (e.g. as recently pointed out by a social-science 
student in a physics class here) seem relevant to processes found in even the 
most complex social systems, including economic systems 
that involve money (a ritually-conserved quantity designed for portability).  
Thus management of energy flows through various forms of available 
work, ways to thermalize that energy reversibly so as to create and preserve 
correlations between and within organisms and their environment, and the 
storage of information in increasingly more replicable forms, are central 
and recurring parts of life's adventure. 

\section{Net surprisal \& the unexpected}
\label{sec:Unexpected}

Entropy, a measure of expected or average {\em surprisal}, has 
been important in thermodynamics since the work of Clausius in 1865, 
although its firm connection to information measures 
is more recent.  {\em Net surprisals}, defined as a difference 
in average surprisals between one state of information and 
another (the second being often some reference or equilibrium 
state), were initially referred to by Gibbs in 1875 as 
availabilities \cite{Knuth66}.  Generalized availabilities, 
negative logarithms of the partition function as shown above, 
are deeply rooted in the mathematics of statistical inference, and 
increasingly recognized for their connection to both free 
energy \cite{Evans69, Tribus71} and complexity \cite{Morowitz68}.
Lloyd's measure of complexity via {\em depth} is in the same 
category \cite{Lloyd89}.

Although we stop with the discussion of net surprisals here, 
it is difficult to resist pointing out that lifeforms, as 
information engines in symbiosis with codes which survive 
by replication, have a vested interest in being able to 
distinguish alternatives with high net surprisal relative to 
an expected ambient (i.e. with respect to what is common).  
The {\em noble} passions ala Fig \ref{SuprisalMap}, e.g. for 
being a good friend/mentor, sibling/parent, citizen/leader, 
believer/cleric, and witness/scholar, may be considered 
evidence of interest in recognizing net surprisal.  

Corollary symptoms of preoccupation with recognizing
high net surprisal include: (a) the positive importance 
in human culture of attributes like special, unique, 
or rare; (b) the human appetite for variety, pleasant 
surprises, and even gambling; (c) the importance of 
special recognition throughout life, including the 
need for attention as a youngster and the need to 
signify, have great ``discretionary power'', or be 
famous as an adult;  (d) the importance of humor, and 
the discovery of novel, fortuitous, and/or surprising 
connections in our language and behavior as a kind of 
``dessert'' for us information processors at the end 
of a long day's work; (e) the use in the vernacular 
of adjectives like ``heated'', ``hot'', and ``steamed'' 
for situations in which action dominates thought 
(e.g. high eV/bit), and adjectives like ``cool'' for 
situations in which information dominates by 
comparison; and (f) of course the desire to see 
the genetic and memetic codes, that we've had a hand 
in designing, fare well with challenges posed by their 
environment in days ahead.

\section{Conclusions}
\label{sec:Conclude}

Statistical physics is perhaps the most quantitative 
tool available for the generalist, in that it allows 
one with meager information to make rigorous assertions 
that a subset of outcomes are going to be impossible 
in practice.  Perpetual motion machines \cite{Penrose70} 
are the classic example.  Although the calculation
methods, and some of the examples used above, are old, 
the problems they can address are contemporary.  

As tools in the ``science of the possible'' these 
methods can be used to show (for example) that 
reversible methods for converting high temperature 
heat to low temperature heat for home heating could 
reduce the energy cost of heating by an order of 
magnitude \cite{Silver81, Jaynes91}, and that going to the store 
to buy a package of automobile seeds is not an 
inconceivable alternative for our descendants a 
century from now \cite{Drexler86}.  Awareness of 
mutual information is crucial to our understanding 
of both quantum computers of the future \cite{Gottesman00}, and 
the molecular machines for replicating nucleic 
acid sequences which keep us going today \cite{Schneider00}.

Concerning the information theory paradigm itself, 
Amnon Katz said in the preface to his 1967 book 
\cite{Katz67} that writing a book on the information 
theory approach to statistical physics was worthwhile 
to him primarily because it provides a coherent 
overview for the novice.  He said that his book 
found little favor with the experts in statistical 
mechanics, because they already knew how to pose 
questions and get answers.  Part of the disinterest 
in a new way to look at things on the part of 
experts \cite{Gopal74} was likely paradigm paralysis 
of the same sort that prompted Swiss watchmakers in 
the late 1960's to discredit as uninteresting, and to 
eventually give away, their own invention of the 
quartz-movement watch along with most of their market 
share in the watchmaking industry \cite{Barker86}.  

Now, a half-century later, the 
pervasive influence of the paradigm shift on mid-level 
physics texts, and the experimental impact of mutual 
information in molecular biology and nano-computing 
research, has left skeptics (even if they are 
legitimately tired of hand-waving metaphors) with little 
to hang onto except the large size of Avogadro's number, 
which makes uncertainty increases associated with heat 
flow tens of orders of magnitude larger than those associated 
with the traditional objects of gambling theory \cite{Lambert99}.
The good news here is that the paradigm shift 
offers additional food for thought to students not 
majoring in physics (especially those involved in the 
code-based sciences).  Of course it is in part the 
responsibilty of physicists to provide such students, 
in an introductory course, with {\em physical} insight into 
quantitative ways for putting these tools to use.

\begin{appendix}
\section{Multiple choice maxent}
\label{app:MultipleChoice}
Suppose we wish to determine the ``state'' of a population of individuals 
with respect to the way they will respond to the questions on an $N$ 
question, $m$ choice multiple-choice test.  The only information that 
we have, however, is the average number of correct answers 
$\kappa \equiv \left< j \right>$ by members of that population.

Let's examine the situation more closely.  There are $m$ ways to answer 
each question, so there are $m^N$ ways to respond to the test as a whole.  
For each possible number of correct answers $j \epsilon \{0,N\}$ there are
\begin{equation}
\text{number of states} = \frac{N!}{j!(N-j)!}(m-1)^{N-j} \equiv d_j \text{.}
\label{StateDensity}
\end{equation}
The function $d_j$ is called the density of states.  Using it, we can 
write the results of the entropy maximization from equations \ref{Solution}, 
\ref{PartitionFn}, and \ref{MaxEnt}, respectively, as
\begin{equation}
p_j = \frac{e^{- \lambda j}}{Z}\text{; }
Z = \sum_{j=0}^{N} d_j e^{-\lambda j}\text{; }
\frac{S}{k} = \ln Z + \lambda \kappa \text{.}
\label{MaxEntMultipleChoice}
\end{equation}

\begin{figure}
\includegraphics{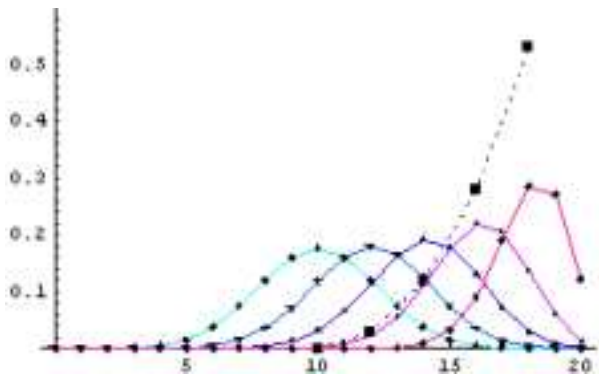}%
\caption{Guessed fraction of students versus 
the number of correct answers out of 20 questions, 
when the average grade for the class is 10, 12, 
14, 16 and 18.  The dashed line is the net 
surprisal (in bits per question) of the 
distribution with given average, relative to that
expected for 20 randomly answered true-false 
questions.}
\label{FigBell}
\end{figure}

The value of $\lambda$ is found by solving the implicit equation 
$\kappa \equiv \sum_{j=0}^{N} j d_j p_j$.  The result 
is $\lambda = - \ln \frac{(m-1) \kappa}{N-\kappa}$, so that 
$Z = {\left(\frac{N(m-1)}{N-\kappa}\right)}^N$, 
$S = k \ln \frac{N^N \left({m-1}\right)^{N-\kappa}}{\kappa^\kappa \left({N-\kappa}\right)^{N-\kappa}}$, 
and
\begin{equation}
d_j p_j = \frac{N!}{j!(N-j)!} {\left(\frac{\kappa}{N}\right)}^j {\left(\frac{N-\kappa}{N}\right)}^{N-j}\text{.}
\end{equation}
When the $d_j p_j$ are plotted as a function of $j$ for a given 
value of $\kappa$ (cf. Fig \ref{FigBell}), one gets the binomial distribution. 
This in turn for large N and values of $\kappa$ not too large or small 
may be approximated by the continuous Gauusian distribution or 
Bell curve.  When $N$ is large but the $\kappa$ is very small by comparison, 
it reduces to the Poisson distribution, useful for predicting random but 
unlikely events like the distribution of meteor impacts.  The net surprisal per 
question (dotted line in Fig \ref{FigBell}) quantifies the amount of mutual 
information about course material (relative to random answers on a true-false 
test) evidenced by students with a given mean score.  This net surprisal measure 
($I_{net} = k N \ln m - S$) can help teachers take into account the fact that a 
score of N/m correct, on an m-choice N-question test, provides zero evidence 
of student learning since random guesses would on average yield the same result.  

The entropy of the state distribution $S$ also quantifies our uncertainty 
about the response of any particular member of the population to the test, 
given only the population average grade.  There is an alternative way 
to look at it as well.  Suppose there was some physical process 
operating which acted only to hold the average 
grade at the specified value, but otherwise had no effect on the 
probability distribution.  If that were the only physical process 
determining the outcome of the test, then we might expect the actual 
spread in test responses to agree with the distribution of responses 
predicted by equation \ref{MaxEntMultipleChoice}, which guesses based 
only the measured value for $\kappa$.  Thus even given
detailed information on the set of responses to the test, it would help 
little in decreasing our uncertainty when trying to predict responses 
because the constraint on $\kappa$ was the only {\em physical} 
process constraining the distribution.

This is more than idle speculation.  Indeed the main control 
variable in educational testing is often the ``difficulty'' of 
the questions, and actual test responses for a given average grade 
do tend to follow the distribution predicted in equation \ref{MaxEntMultipleChoice}.
Failure of this to occur is thus evidence for other constraints 
operating (e.g. the presence of two populations of students).  By 
the same token, agreement between the predicted distribution and the 
actual one cannot, however, be taken as evidence that no other 
physical principles are active in the system, only that the results 
of the test probably tell us little about them.

In the foregoing example, the response of a population sample to a
test was classified into a set of ``microstates''.  The uncertainty 
about the response given only information about the average grade 
was shown to agree with the spread in measured values, from which it 
was inferred that the major physical constraint acting on the response 
distribution was in effect one which determined the average value.  
In a more general sense when we consider all of the physical microstates 
accessible to a given system, the physical entropy of that system 
is defined as the uncertainty calculated when all externally 
detectable constraints on the state distribution are used as 
constraints in the maxent calculation used.  It is thus the 
minimum uncertainty possible, based on all of the ``mutual information'' 
about the system available to the world outside.  

\section{Mutual information basics}
\label{app:MutualInformation}

A discussion of correlated subsystems which, between 
themselves, house mutual information must begin with a 
definition of subsystems.  Such subsystems 
are variously defined, for example, as individual particles, 
as collections of particles, as individual states 
(which may or may not be occupied with particles), and 
as regions or control volumes in and out of which energy 
and mass might flow.  Our earlier distinction between a
steady state engine and its environment, as well as 
the molecule, cell, tissue, metazoan, gene-pool and 
meme-pool boundaries discussed in section 
\ref{sec:CodesExcitations}, also fall in this category.

Even the isolated-system second law itself must first 
separate the world into observed-system and observer, 
because as 
we mentioned earlier it also is an assertion about the 
time evolution of mutual information, namely:  The 
correlation information that an observed physical system 
has with an environment from which it is isolated will 
not decrease with time, and will likely increase since 
information available to the environment needed to 
model propagation of the isolated system through time 
may fall short.

Once some hopefully useful boundaries for the subsystems 
of interest in a given problem have been defined, a set 
of $N_{ss}$-subsystem joint probabilities can be defined.  
For $N_{ss}=2$, {\em joint probabilities} $p[ij]\geq 0$ 
obey $\sum_i \sum_j p[ij] = 1$.  Here the $i$ indices 
run over all possible states for subsystem I, while the $j$ 
indices run over all possible states for subsystem J.  
From the joint probabilities one can calculate 
{\em marginal probabilities} like $p[i]\equiv \sum_j p[ij]$, 
which ignore the state of other subsystems, and 
{\em conditional probabilities} like 
$p_i[j]\equiv \frac{p[ij]}{p[i]}$ associated with a 
specific state of one subsystem (here the ith state 
of subsystem I).  From these probabilities then values for
{\em joint entropy} $S[IJ]/k\equiv \left<-\ln p[ij] \right>$, 
{\em marginal entropies} like $S[I]/k\equiv \left<-\ln p[i] \right>$ 
and {\em conditional entropies} like $S_I[J]/k\equiv \left<-\ln p_i[j] \right>$
follow immediately.  Mutual or correlation information between 
systems I and J, denoted here as $M[I,J]$ and defined by 
equation \ref{mutual_inf} as the sum of marginal entropies 
$S[I]+S[J]$ minus the joint entropy $S[IJ]$, thus becomes
\begin{equation}
M[I,J] = - k \sum_i{\sum_j p[ij]}\ln\frac{p[i]p[j]}{p[ij]}\geq 0\text{.}
\label{Mutual2}
\end{equation}
Thus from equation \ref{NetSuprisal} it appears that mutual 
information is simply the net surprisal that follows upon
learning that systems (here I and J) are not independent.

Examples of correlated subsystem pairs include 
photon or electron pairs with opposite but 
unknown spins, a single strand of messenger RNA 
and the sequence of nucleotides in the gene from 
which it was copied, a manuscript and a copy of 
that manuscript created with a xerox machine (or 
a video camera), your understanding of a subject 
before being given a test and the answer key 
used by the teacher to grade that test (hopefully), 
enzymes and coenzymes with site specificity, tissue 
sets treated as friendly by your immune system, 
metazoans who developed from the same genetic 
blueprints (e.g. identical twins), families that 
share similar values, and cities which occupy 
similar niches in different cultures (e.g. 
sister cities).  As you might imagine this 
list of subsystem pairs is incomplete, and 
many of the quantities listed remain difficult 
to quantify. 

With increasing $N_{ss}$, the number of marginal and 
conditional entropies increases rapidly, and many 
new mutual information terms (all positive) emerge 
as well.  For example, when $N_{ss}=3$, marginal 
probabilities exist which ignore either one or two 
sub-systems (e.g. like $p[ij]\equiv \sum_k p[ijk]$ and 
$p[i]\equiv \sum_j \sum_k p[ijk]$).  Similarly conditional 
probabilities can specify the state of one or two sub-systems.  
These all give rise to analogous marginal and conditional 
entropies.  Lastly, a set of seven mutual information terms 
can be calculated:  
the joint correlation $M[I,J,K]\equiv S[I]+S[J]+S[K]-S[IJK]$, 
three one-on-two terms like $M[I,JK]\equiv S[I]+S[JK]-S[IJK]$, and 
three one-on-one terms like $M[I,J]\equiv S[I]+S[J]-S[IJ]$. 

These mutual information terms all have positive values which 
are independent of argument ordering, e.g. $M[I,J]=M[J,I]$.  
One useful identity is $M[I,J,K]=M[I,JK]+M[J,K]$, which 
in words says that ``the joint mutual information of systems 
I, J and K is the correlation information between system I 
and system JK, plus that between systems J and K''.  Another 
relationship that we conjecture here is $M[I,JK]\geq M[I,J]$, 
or in other words:``System I and system JK have at least as 
much in common as do systems I and J alone''.  

The maxent formalism, of course, automatically estimates 
joint probabilities, from which all of these quantities 
follow.  Figuring out how to constrain the maximization 
with knowledge of mutual information between subsystems is 
therefore the primary challenge in adapting it to such 
problems.

 



\end{appendix}
\begin{acknowledgments}
I would like to thank the late E. T. Jaynes for many interesting papers and 
discussions over the past half century, Tom Schneider at NIH for spirited 
questions and comments, and other colleages in the regional and national AAPT 
content modernization communities too numerous to name.
\end{acknowledgments}


\bibliography{ifzx.bib}

\end{document}